\documentstyle[seceq,wrapfig]{ptptex}
\title{%
Flavor and Horizontal Symmetries}

\author{%
Paul H. {\sc Frampton}\footnote{Supported in part by the U. S. Department
of Energy.}
}

\inst{%
Institute of Field Physics, Department of Physics and Astronomy,
University of North Carolina, Chapel Hill, NC 27599-3255, USA
}

\abst{%
After a brief introduction to what are the basic flavor questions to be
addressed,
I introduce the underlying ideas of horizontal symmetries, with group $G_H$.
For the purposes of specific model building, it is useful to
classify models according to the scale at which $G_H$ is broken.
I consider the three cases: below $m_t$; somewhat above $m_t$; and at
$M_{GUT}$.
After a discussion of the shadow sector in the $E_6 \times E_8'$ superstring,
there is a summary.
}
\begin{document}

\maketitle

\section{Flavor Questions.}
We may identify three issues:

{\it (i)Replication.}

Surely the most basic flavor question is why there exists the replication of
the quarks: u,d; c,s; t.b and similarly for the leptons: $\nu_e,e;
\nu_{\mu},\mu;
\nu_{\tau},\tau$.

Why are there three families?  Although there are papers about this topic, it
will not be
my subject here.

{\it (ii)Fermion mass hierarchy.}

If I define $\lambda = sin\theta_C \simeq 0.22$, the Cabibbo angle, as a
"small" parameter
then for the up-type (Q=+2/3) quarks:
\begin{equation}
\ m_u/m_t \sim \lambda^8; ~~~~~~~~~~   m_c/m_t \sim \lambda^4
\end{equation}

\noindent
while for the down-type quarks (Q = -1/3):
\begin{equation}
m_d/m_b \sim \lambda^4; ~~~~~~~~~~   m_s/m_b \sim \lambda^2.
\end{equation}

\noindent
The charged lepton masses approximately satisfy:
\begin{equation}
m_e/m_{\tau} \sim \lambda^4; ~~~~~~~~~   m_{\mu}/m_{\tau} \sim \lambda^2.
\end{equation}

\noindent
These masses are evaluated at the GUT scale $\sim 2 \times 10^{16}$GeV.
Whence do such hierarchies arise?

{\it (iii)Mixing Hierarchy.}

When the quark mass matrices are diagonalized with the
usual bi-unitary transformations:

\begin{equation}
U_LM^UU_R^{\dagger} = diag(m_u, m_c, m_t) ~~~~~~~~~~
D_LM^DD_R^{\dagger} = diag(m_d, m_s, m_b) ~~~~~~~~~~
\end{equation}
and the CKM matrix is constucted by $V_{CKM} = U_LD_L^{\dagger}$ one
find that its elements have the hierarchy:

\begin{equation}
  |V_{ii}\sim1| > |V_{12}| > |V_{23}| > |V_{13}|
\end{equation}

\noindent
where these four elements are of order $1, \lambda, \lambda^2, \lambda^3$
respectively.
Whence do these hierarchies come? ......is the third and last flavor question.

Note that the fermion and mixing hierarchies speak to 12 of the 19 parameters
of the Standard Model: the $m_i$ and the $\Theta_k$. The others 7 are: the
three couplings $\alpha_i$; the two CP parameters $\delta$ and
$\overline{\theta}$;
and the two scales $M_W$ and $M_H$ arising in the electroweak symmetry
breaking.

\section{Horizontal Symmetries}

Let the horizontal symmetry group be $G_H$. There are choices to be made about
$G_H$:
whether it is global or gauged, finite or infinite(Lie), abelian or
non-abelian.

My choices will be: gauged, finite and non-abelian.

Finite abelian groups ($Z_N$) \cite{Ibanez} and finite non-abelian groups
($S_N$ only)
\cite{Sugawara} have been studied previously.

What I mean by a gauged finite group will be discussed below.

First I offer a brief review of all finite groups of order $g \leq 31$. [One
usually
stops at $g = 2^n - 1$ because of the richness of $g = 2^n$] In this range
there are,
according to standard textbooks\cite{FKe2}, 93 inequivalent groups of which 48
are
abelian and 45 are non-abelian. $g \geq 32$ might be interesting too but lower
g is surely
more economic.

Let me deal quickly with all the abelian cases. The building block is $Z_p$
where the elements are
the $p^{th}$ roots of unity, $e^{2 \pi i/p}$. The only fact one needs is that
$Z_p \times Z_q$ is equivalent to $Z_{pq}$ if and only if p,q have
no common prime factor. So if I decompose the order g of the group into its
prime factors
$g = \prod_i p_i^{k_i}$ then the number of inequivalent abelian finite groups
at order g is $N_a(g) = \prod_{k_i} P(k_i)$, where $P(\nu)$ is the number
of ordered partitions of $\nu$. For example, $P(1,2,3,4) = 1,2,3,5$.

For the cases $g \leq 31$, one finds $N_a(g) = 1$ {\it except} that $N_a(g) =
2$
for g = 4,9,12,18,20,25,28; $N_a(g) = 3$ for 18,24,27; and $N_a(16) =5$.

Thus, adding these results gives the required answer of 48 abelian groups
with $g \leq 31$. These will not be considered further here.

Now, and for the remainder of the talk, I shall consider non-abelian finite
groups.

The best known are the $S_N$ permutation (or symmetric) groups with order g =
N!
Since g grows so rapidly, only N=3,4 are in our range. Generally, $S_N \subset
O(N - 1)$ and can be interpreted geometrically as the symmetry of a regular
N-plex in $( N -1 )$ spatial dimensions.

Next come the dihedral groups $D_N$, with order g = 2N, which are subgroups of
0(3).
The geometrical interpretation is the symmetry of a 2-sided planar regular
N-agon
in 3 spatial dimensions where the polygon is treated as a "two-faced" entity.

There is the one g = 12 tetrahedral group (T) which is the even-permutation
subgroup
of $S_4$.

Spinorial generalizations (doubles) of the $D_N$ are the dicyclic groups
$Q_{2N}$
which have order g = 4N and are subgroups of SU(2) rather than O(3).

The majority (32) of the non-abelian groups with order $g \leq 31$ are made
from $S_N, D_N, T, Q_{2N}$
as follows:

g = 6:  $D_3 ( \equiv S_3 )$.

g = 8:  $ D_4; Q \equiv Q_4$.

g = 10: $ D_5 $.

g = 12: $D_6; Q_6; T$.

g = 14: $D_7$.

g = 16: $D_8; Q_8; Z_2 \times D_4; Z_2 \times Q$.

g = 18: $D_9; Z_3 \times D_3$.

g = 20: $D_{10}; Q_{10}$.

g = 22: $D_{11}$.

g = 24: $D_{12}; Q_{12}; Z_2 \times D_6; Z_2 \times Q_6; Z_2 \times T; Z_3
\times D_4; D_3 \times Q; Z_4 \times D_3; S_4$.

g = 26: $D_{13}$.

g = 28: $D_{14}; Q_{14}$.

g = 30: $ D_{15}; D_5 \times Z_3; D_3 \times Z_5$.
There are another 13 which are twisted products of $Z_n$:

g = 16: $Z_2 \tilde{\times} Z_8 (two); Z_4 \tilde{\times} Z_4; Z_2
\tilde{\times} (Z_2 \times Z_4) (two) $.

g = 18: $Z_2 \tilde{\times} (Z_3 \times Z_3)$.

g = 20: $Z_4 \tilde{\times} Z_5$.

g = 21: $Z_3 \tilde{\times} Z_7$.

g = 24: $Z_3 \tilde{\times} Z_q; Z_3 \tilde{\times} Z_8; Z_3 \tilde{\times}
D_4$.

g = 27: $Z_3 \tilde{\times} Z_9; Z_3 \tilde{\times} (Z_3 \times Z_3)$.

Of these groups $Q_{2N}$ is of most interest to
model-building\cite{FKe1,FKe2,FKo1,FKo2}.

The group $Q_{2N}$ with order g = 4N has 4 singlet representations $1$,
$1^{'}$, $1^{''}$,$ 1^{'''}$
and the $(N - 1)$ doublets $2_{(j)}, 1 \leq j \leq (N-1)$. The doublet
multiplication is:

\begin{equation}
 2_{(j)} \times 2_{(k)} = 2_{|j-k|} + 2_{min(j+k,N-j-k)}
\end{equation}
with the generalized notation:

\begin{equation}
 2_{(0)} \equiv 1 + 1^{'} ~~~~~~~~~~~~~~  2_{(N)} \equiv 1^{''} + 1^{'''}
\end{equation}
$Q_{2N}$ has only singlet and doublet reprentations.

To obtain a clearer intuitive understanding of $Q_{2N}$, it is defined by the
algebra:

\[ A^{2N} = E ~~~~~ B^2 = A^N ~~~~~ ABA = B \]
The elements are

\[ E, A, A^2, A^3, ........A^{2N-1}, B, AB, A^2B, A^3B,.........A^{2N-1}B \]
A simple matrix representation is:

\[ A = \left( \begin{array}{cc} cos\theta & sin\theta \\ -sin\theta & cos\theta
\end{array} \right)
\]
with $\theta = \pi/N$ because then

\[ A = \left( \begin{array}{cc} cosN\theta & sinN\theta \\ -sinN\theta &
cosN\theta \end{array} \right)
 = \left( \begin{array}{cc} -1 &   \\  & -1 \end{array} \right) \]
and for B I use the simple matrix

\[ B = \left( \begin{array}{cc} -i &   \\  & -i \end{array} \right)  \]
$Q_{2N}$ is thus the full SU(2) symmetry of a planar regular N-agon where
rotation by $2\pi$ gives a sign -1; only rotation by $4\pi$ brings back the
identity.

\subsection{Gauging $G_H = Q_{2N}$}

Gauging $G_H$, a finite group, is subtle as is immediately seen by considering
the
covariant derivative. In fact, at a local level it is meaningless in a flat
spacetime neighborhood.
Globally, with respect to topological aspects of the spacetime manifold, it is
best
done by gauging $SU(2)_H \supset Q_{2N}$ then spontaneously breaking to
$Q_{2N}$. If
this is at a high scale, the effective theory has no gauge field - but
consistency with wormholes is preserved.

Gauging leads to consistency requirements:

(a) Chiral fermions must be in complete representations of $SU(2)_H$.

(b) $(SU(2)_H)^2Y$ anomalies must cancel.

(c) Witten's global $SU(2)_H$ anomaly must cancel.

(b) and (c) are straightforward but (a) requires the dictionary for embedding
$Q_{2N} \subset SU(2)$.
This is actually a simple pattern:

\[ SU(2) \rightarrow Q_{2N} \]

\[ 1 \rightarrow 1 \]

\[ 2 \rightarrow 2_1 \]

\[ 3 \rightarrow 1^{'} + 2_2 \]

\[4 \rightarrow 2_1 + 2_3 \]

\[ 5 \rightarrow 1 + 2_2 + 2_4 \]
and so on. The infinite sequence is clear from the above.

\section{ Model Building with Horizontal $Q_{2N}$ Symmetry}

There are five "triples" in the standard model for which $Q_{2N}$ assignments
must be
made:

(1)  $ (t, b )_L, (c ,s )_L, (u , d )_L$.

(2)  $ t_R, c_R. u_R$.

(3)  $ b_R, s_R, d_R$.

(4)  $ (\nu_{\tau}, \tau)_L, (\nu_{\mu} ,\mu )_L, ( \nu_e ,e )_L$.

(5)  $ \tau_R, \mu_R, e_R$.

These must be assigned to anomaly-free complete representations of $SU(2)_H
\supset G_H \equiv Q_{2N}$.
The technical details depend on the $G_H$ breaking scale. I shall consider: A
10 GeV, B 10 TeV, C $10^{16}$ GeV.

\subsection{$\Lambda_H \sim 10GeV (> m_b)$}

Assignments are arranged around the top quark such that:

t mass is a $G_H$ singlet.

$b, \tau$ masses break $G_H \rightarrow G^{'}$.

c mass breaks $G^{'} \rightarrow G^{''}$

$s, \mu$ masses break $G^{''} \rightarrow G^{'''}$

At the same time, I exclude, or minimize, additional fermions and demand full
anomaly cancellation.

It is possible to show\cite{FKe2} that of all the 93 finite groups with $g \leq
31$, only the dicyclic
groups remain as candidates for $G_H$.
The simplest model uses $Q_6$. Recall that:

\[ 1 \rightarrow 1  \]

\[ 2 \rightarrow 2_{(1)} \]

\[ 3 \rightarrow 1^{'} + 2_{(2)} \]
so for the five triples the only possible assignments are:

\[ 1 + 1 = 1 \]

\[ 1^{'} + 2_{(2)} \]

\[ 1 + 2_{(1)} \]

For the $Q_6$ model, the assignments are:

\[\begin{array}{ccccccc}

\left( \begin{array}{c} t \\ b \end{array} \right)_{L} &
1 & \begin{array}{c} t_{R}~~~ 1 \\ b_{R} \hspace{0.2in}1^{'} \end{array} &
\left( \begin{array}{c} \nu_{\tau} \\ \tau \end{array} \right)_{L} & 1 &
\tau_{R} & 1^{'} \\

\left. \begin{array}{c} \left( \begin{array}{c} c \\ s \end{array} \right)_{L}
\\
\left( \begin{array}{c} u \\ d \end{array} \right)_{L} \end{array}  \right\} &
2_{S}
 &  \begin{array}{c} \left. \begin{array}{c} c_{R} ~~~ 1\\ u_{R} ~~~ 1
\end{array} \right. \\
\left. \begin{array}{c} s_{R} \\ d_{R} \end{array} \right\} 2 \end{array} &
\left. \begin{array}{c} \left( \begin{array}{c} \nu_{\mu} \\ \mu \end{array}
\right)_{L} \\
\left( \begin{array}{c} \nu_{e} \\ e \end{array} \right)_{L} \end{array}
\right\} & 2_{S}
& \left. \begin{array}{c} \mu_{R} \\ \\ e_{R} \end{array} \right\} & 2

\end{array} \]
The mass matrix textures are:
\[  U = \left( \begin{array}{cc} <2_S> &  <2_S>  \\ <1> & <1> \end{array}
\right)   \]

and:

\[D = L = \left( \begin{array}{cc} <1''+1'''+2_S> & <2_S> \\  <2> & <1'>
\end{array} \right) \]

The symmetry-breaking steps involve a VEV to a $Q_6$ singlet $\Rightarrow t$
mass; then
 a VEV to a $1^{'}$ breaks $Q_6$ to $Z_6$, providing $b, \tau$ masses.
A $(1,2_{(1)}) VEV$ gives the c mass and finally a $(2, 1^{''} + 1^{'''})$ VEV
gives
the $s, \mu$ masses.

Some remarks on the $Q_6$ model:

(i) The $(SU(2)_H)^2Y$ anomaly cancellation requires certain extra singlets
predicted to lie between 50GeV and 200GeV.

(ii) The hierarchy of Yukawa couplings has been removed: they now all lie
between 0.1 and 1.0.

(iii) It provides a first step to understanding why the top quark mass
is so different from the other quark masses.

\subsection{Breaking of $G_H$ at $\geq 1TeV$}

Here we shall use non-abelian horizontal symmetry
in connection with derivation of an ansatz for the texture zeros
in the quark mass matrices.

The horizontal symmetry will be again $Q_{2N}$, but now the Froggatt-Nielsen
mechanism\cite{FN} becomes an essential part of mass generation. This means
that additional vector-like pairs of fermions, at high scale, are present
- and these must likewise be assigned to representations
of $G_H$.

The quark assignments are:

\[  \begin{array}{ccc}

\left.
\begin{array}{c} \left( \begin{array}{c}  t \\ b \end{array} \right)_{L} \\
\left( \begin{array}{c}  c \\ s \end{array} \right)_{L} \end{array}
\right\} &
2_{(2)} &
\left.
\begin{array}{c} \left. \begin{array}{c} t_R \\ c_R \end{array} \right\}
 ~~~2_{(2)} \\  \left. \begin{array}{c}  b_R \\ s_R \end{array} \right\}~~~
2_{(1)} \end{array} \right.    \\

\left( \begin{array}{c} u \\ d \end{array} \right)_{L} & 1^{'} & \left.
\begin{array}{c} u_R~~~~~ 1^{'}
\\ d_R ~~~~~1 \end{array} \right.  \\

\end{array} \]

The lepton doublets are correspondingly $( 2_{(1)} + 1 )$ and the singlets
$(2_{(2)} + 1^{'})$.
This assignment is completely anomaly-free.
To consider the quark masses I put the SM Higgs doublet in
to down quarks. Their VEVs give masses only to the third family.
The other elements of the up-quark mass matrix $M_U$ arise from
Froggatt-Nielsen graphs shown in Fig. 1.

%\documentstyle[preprint,prl,aps]{revtex}

%\newcommand{\CALL}{{\cal L}}
%\newcommand{\CALH}{$\cal H$}
%\newcommand{\gsim}{\stackrel{>}{\sim}}
%\newcommand{\lsim}{\stackrel{<}{\sim}}
%\newcommand{\aqu}{\langle Q^2 \rangle}
%\newcommand\GeV{\,\mbox{GeV}}
%\newcommand\TeV{\,\mbox{TeV}}
%\newcommand\pB{[\mbox{pb}]}
%\newcommand\SM{$SU(3)_c~\times~SU(2)_L~\times~U(1)_Y$}
%\newcommand\MSbar{$\overline{\mbox{MS}}$}
%\newcommand\order{{\cal O}}

%\begin{document}

%\newpage

%\begin{flushleft}
\begin{figure}[h]

\vspace*{2.0cm}

\setlength{\unitlength}{1.0cm}

\begin{picture}(15,15)

\thicklines

\put(1,13.5){\framebox(2.3,1){$(a)$: $(M_u)_{32}$}}
\put(7.08,12){\vector(1,0){1}}
\put(8.08,12){\line(1,0){0.84}}
\multiput(9,12)(0,0.3){9}{\line(0,1){0.25}}
\put(8.85,14.6){${\bf \times}$}
\put(9.08,12){\line(1,0){0.84}}
\put(10.92,12){\vector(-1,0){1}}
\multiput(11,12)(0,0.3){9}{\line(0,1){0.25}}
\put(10.85,14.6){${\bf \times}$}
\put(11.08,12){\line(1,0){0.84}}
\put(12.92,12){\vector(-1,0){1}}
\put(6.5,11.9){$t_L$}
\put(9.2,14.6){$ \left \langle {H_u} \right \rangle $}
\put(10.2,11.5){$U_{6}$}
\put(11.2,14.6){$ \left \langle {2_8} \right \rangle $}
\put(13,11.9){$c_R$}

\put(1,8.5){\framebox(2.3,1){$(b)$: $(M_u)_{22}$}}
\put(7.08,7){\vector(1,0){1}}
\put(8.08,7){\line(1,0){0.84}}
\multiput(9,7)(0,0.3){9}{\line(0,1){0.25}}
\put(8.85,9.6){${\bf \times}$}
\put(9.08,7){\line(1,0){0.84}}
\put(10.92,7){\vector(-1,0){1}}
\multiput(11,7)(0,0.3){9}{\line(0,1){0.25}}
\put(10.85,9.6){${\bf \times}$}
\put(11.08,7){\line(1,0){0.84}}
\put(12.92,7){\vector(-1,0){1}}
\put(4.5,6.9){$c_L$}
\put(9.2,9.6){$ \left \langle {H_u} \right \rangle $}
\put(10.2,6.5){$U_{6}$}
\put(11.2,9.6){$ \left \langle {2_8} \right \rangle $}
\put(13,6.9){$c_R$}
\put(5.08,7){\vector(1,0){1}}
\put(6.08,7){\line(1,0){0.84}}
\multiput(7,7)(0,0.3){9}{\line(0,1){0.25}}
\put(6.85,9.6){${\bf \times}$}
\put(7.2,9.6){$ \left \langle {2_8} \right \rangle $}
\put(7.5,6.5){$Q_{10}$}

\put(1,3.5){\framebox(2.3,1){$(c)$: $(M_u)_{13}$}}
\put(7.08,2){\vector(1,0){1}}
\put(8.08,2){\line(1,0){0.84}}
\multiput(9,2)(0,0.3){9}{\line(0,1){0.25}}
\put(8.85,4.6){${\bf \times}$}
\put(9.08,2){\line(1,0){0.84}}
\put(10.92,2){\vector(-1,0){1}}
\multiput(11,2)(0,0.3){9}{\line(0,1){0.25}}
\put(10.85,4.6){${\bf \times}$}
\put(11.08,2){\line(1,0){0.84}}
\put(12.92,2){\vector(-1,0){1}}
\multiput(13,2)(0,0.3){9}{\line(0,1){0.25}}
\put(12.85,4.6){${\bf \times}$}
\put(13.08,2){\line(1,0){0.84}}
\put(14.92,2){\vector(-1,0){1}}
\put(6.5,1.9){$t_L$}
\put(9.2,4.6){$ \left \langle {H_u} \right \rangle $}
\put(10.2,1.5){$U_{6}$}
\put(11.2,4.6){$ \left \langle {2_8} \right \rangle $}
\put(15,1.9){$u_R$}
\put(12.2,1.5){$U_{14}$}
\put(13.2,4.6){$ \left \langle {2_{14}} \right \rangle $}

\end{picture}

\caption{Froggatt-Nielsen tree graphs for $M_u$. (The symmetric counterparts,
$(M_u)_{23},$ another graph for $(M_u)_{22}$ and $(M_u)_{31}$ are not shown)}

\label{Fig. 1}

\end{figure}

%\end{flushleft}

%\end{document}

There are second graphs, not shown in Fig.1, for $(M_U)_{32}$, $(M_U)_{31}$.
Similar graphs for $(M_D)$ lead to the textures:

\begin{equation}
M_U = \left( \begin{array}{ccc} 0 & 0 & \lambda^4 \\ 0 & \lambda^4 & \lambda^2
\\
\lambda^4 & \lambda^2 & 1  \end{array} \right)
\end{equation}

\begin{equation}
M_D = \left( \begin{array}{ccc} 0 &  \lambda^4  & 0 \\  \lambda^4  &  \lambda^3
 & 0 \\
0 & 0 & 1   \end{array} \right)
\end{equation}

Here, as before, $\lambda \simeq sin\theta_C$. The expansion parameter is also
identified with the ratios:

\begin{equation}
\lambda = <S_i>/M_{odd}    ~~~~~~~~~\lambda^2 = <S_i>/M_{even}
\end{equation}
where the "odd" $Q_6$ doublets occur in the $SU(2)_H$ spinor representions ;
the "even"
in the vector ones.

Another symmetric texture (the only other attainable one with five zeros) is:

\begin{equation}
M_U = \left( \begin{array}{ccc} 0 & \lambda^6 &  0 \\ \lambda^6 & \lambda^4 &
\lambda^2 \\
0 & \lambda^2 & 1  \end{array} \right)
\end{equation}

\begin{equation}
M_D = \left( \begin{array}{ccc} 0 &  \lambda^4  & 0 \\  \lambda^4  &  \lambda^3
 & 0 \\
0 & 0 & 1   \end{array} \right)
\end{equation}
Three other phenomenologically-viable textures\cite{RRR} which are symmetric
with five zeros
are not attainable and hence disfavored.

The postulation of zeros in the mass matrices reduces the number of free
parameters
in the low energy theory. This now has a dual description in terms of a
horizontal symmetry $Q_{2N} \subset SU(2)_H$.

This $SU(2)_H$ could arise from a GUT group or directly from a superstring.

My main point is that derivation of the values of the masses
in a putative theory of everything may likely involve a horizontal symmetry,
probably
gauged, as an important intermediate step. The two simple cases given above
illustrate how this can
happen.

\subsection{$G_H$ Breaking $\geq M_{GUT}$}

Again the Froggatt-Nielsen mechanism is used, invoking a spectrum of superheavy
fermions
in vector-like pairs.

The symmetry group\cite{FKo2} is $SU(5) \times SU(5) \times SU(2)_H$ The
$SU(2)_H$ is broken
to $Q_{12}$ at M(SU(2)) and the $SU(5) \times SU(5)$ is broken to a diagonal
$SU(5)$
at $M_{GUT}$. Supersymmetry is broken near the weak scake.

The light, heavy and superheavy fermion contents are in the following list
of chiral $SU(5) \times SU(5)$ supermultiplets. The third entry denotes
the content under $S(2)_H \rightarrow Q_{12}$.

\[ (10, 1, 1 \rightarrow 1(T)) \]

\[ (10, 1, 4 \rightarrow 2_1+2_3 ) \]

\[ (10, 1, 7 \rightarrow 1^{'}+2_2+2_4+2_6) \]

\[ (\overline{10}, 1, 4 \rightarrow 2_1+2_3) \]

\[ (\overline{10}, 1, 7 \rightarrow 1^{'}+2_2+2_4+2_6) \]

\[ (\overline{5}, 1, 3 \rightarrow 1^{'}+2_2) \]

\[ (\overline{5}, 1, 6 \rightarrow 2_1+2_3+2_5(H_d/b) ) \]

\[ (5, 1, 1(H_u) ) \]

\[ (5, 1, 3 \rightarrow 1^{'}+2_2 ) \]

\[ (5 , 1, 4 \rightarrow 2_1+2_3) \]

\[ (1 , 10, 1) \]

\[ (1 , 10, 2 \rightarrow 2_1(Q) ) \]

\[ (1 , \overline{10}, 1) \]

\[ (1 , \overline{5}, 2 \rightarrow 2_1(D) ) \]

This list is seen to be relatively short when one realizes that it includes all
the
vector-like F-N superheavy fermions and the light chiral fermions.

The effective theory at the weak scale contains only the fermions of the
standard
model transforming as follows: $(u,d)_L, (c,s)_L$ as $Q(2_1)$; $(t,b)_L$ as
T(1);
$u_R,c_R$ as $U(2_1)$; $d_R,s_R$ as $D(2_1)$; $t_R$ as T(1);  $b_R$ as $H_d/b
(2_5)$.

This gives supersymmetry without R-parity. $b-\tau/H_d$ are in a horizontal
doublet
incompatible with the usual R-parity, but matter parity arises here from group
properties of
$G_H$. The "$\mu$ problem" ($\overline{5}.5$ term at tree level) is also solved
here by $G_H$.

The mass matrix textures are:

\begin{equation}
M_U = \left( \begin{array}{ccc} \lambda^8 & \lambda^6 &  0 \\ \lambda^6 &
\lambda^4 & 0
 \\
0 & 0 & 1  \end{array} \right)
\end{equation}

\begin{equation}
M_D = \left( \begin{array}{ccc} 0 &  \lambda^5  & \lambda^5 \\  \lambda^4  &
\lambda^3  &
\lambda^3 \\
0 & \lambda^3 & 1   \end{array} \right)
\end{equation}
This is a phenomenologicaly viable SusyGUT for the low energy parameters.

\section{Shadow $E_8^{'}$ Sector}

For the heterotic string, the $E_8 \times E_8^{'}$ becomes on Calabi-Yau
compactification typically $E_6 \times E_8^{'}$ and the $E_6$ can become
$SU(3)^3$,
for example. In the visible sector one then has unification at $M_{GUT} \sim 2
\times 10^{16} GeV$ with $\alpha_{GUT}^{-1}\sim 25$.

The $SU(3)^3$ couplings have $\beta = 0$ and consistency dictates an
$M_{string} =
3.5 \times 10^{17} GeV$. Bridging across to the shadow sector one can choose
an $SU(N)$ gauge subgroup of $E_8^{'}$ such that $\alpha_N$ becomes O(1)
at a scale where a gluino condensate may break supersymmetry. This suggests
{\it e.g.}
$SU(5) \times SU(4) \times U(1)$. It is possible that the shadow photino can
act as
cosmological dark matter\cite{FKW}.

\section{Summary}

The flavor questions for fermion mass and mixing hierarchies may require
gauged horizontal symmetries and the above examples illustrate how the dicyclic
groups are well-suited. The covering $SU(2)_H$ is gauged for consistency.
The cases I have described show how this can  happen and how it can give
phenomenologically acceptable mass matrices and mixings.

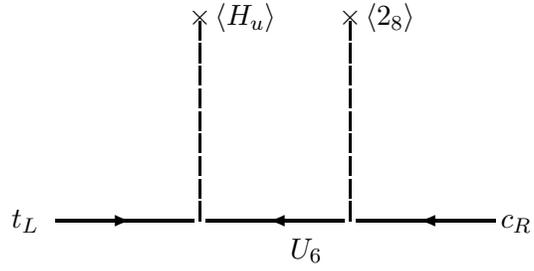
\end{document}